# Towards advanced-fuel fusion:

# Electron, ion energy >100keV

# in a dense plasma

Eric J. Lerner

*Lawrenceville Plasma Physics, 9 Tower Place, Lawrenceville, NJ 08648*

**Controlled fusion with advanced fuels requires average electron and ion energies above 100 keV (equivalent to 1.1 billion K) in a dense plasma. We have met this requirement and demonstrated electron and ion energies over 100 keV in a compact and inexpensive dense plasma focus device. We have achieved this in plasma "hot spots" or plasmoids that, in our best results, had a density-confinement-time-energy product of $5.0 \times 10^{15}$ keVsec/cm$^3$, a record for any fusion experiment. We measured the electron energies with an X-ray detector instrument that demonstrated conclusively that the hard X-rays were generated by the hot spots.**

Controlled fusion with advanced fuels, especially hydrogen-boron-11, is an extremely attractive potential energy source. Such fuels generate nearly all their energy in the form of charged particles, not neutrons, thus minimizing or eliminating induced radioactivity. They also allow direct conversion of charged- particle energy to electric power, without the expensive intermediate step of generating steam for turbines.[1-3] However, fusion with such fuels requires average electron and ion energies above 100 keV (equivalent to 1.1 billion K) in a dense plasma. We have met this requirement and demonstrated electron and ion energies over 100 keV in a dense plasma focus device which is compact and inexpensive. We have achieved this in plasma "hot spots" or plasmoids that, in our best results, had a density-confinement-time-energy product of $5.0 \times 10^{15}$ keVsec/cm$^3$, a record for any fusion experiment and a factor of ten above the best achieved in the much larger Tokamak experiments.[4] To do this, we used a higher-than-usual operating pressure for the device, which increased particle energy. We measured the electron energies with an X-ray detector instrument that eliminated previously-existing doubts that the hard X-rays were generated by the hot spots.

The dense plasma focus (DPF), first invented in 1954, is far more compact and economical than other controlled-fusion devices. It consists of two coaxial cylindrical electrodes usually less than 30 cm in all dimensions in a gas-filled vacuum chamber connected to a capacitor bank. The total cost of such a device is under $500,000, for a 500 kJ system. It is capable of producing high-energy X-ray and gamma-ray radiation and intense beams of electrons and ions, as well as abundant fusion reactions.[5] In operation, the capacitors discharge in a several-microsecond pulse, the gas is ionized and a current sheath, consisting of pinched current filaments, forms and runs down the electrodes. When the sheath reaches the end of the inner electrode (the anode), the

filaments pinch together, forming dense, magnetically-confined, hot spots or plasmoids.[6-8] The plasmoids emit soft X-rays with energy in the range of several kiloelectron volts. X-ray pinhole images have demonstrated that the plasmoids are tiny, with radii of a few microns to tens of microns.[9-10] The plasmoids have densities in the range of $10^{20}$ - $10^{21}$ /cm$^3$. These densities have been measured in a number of independent methods including heavy ion fusion[11], $CO_2$ laser scattering,[12] and x-ray line intensities[13].

These plasmoids emit intense beams of accelerated ions and electrons.[14-15] The electron beam transfers part of its energy to the plasmoid electrons, which generate X-rays through collisions with nuclei. Through a plasma instability, the electrons then transfer part of their energy to the ions, with a typical delay (in our experiments) of ~10 ns. Ion collisions, generating fusion reactions and neutrons, then occur through the intersection of trapped beams.[16] When the ion and electron beams have exhausted the magnetic energy that confines the plasmoid, and partially or wholly evacuated the particles in the plasmoid, the fusion reactions end.

The DPF routinely produces hard X-rays and gamma rays indicating the presence of bremsstrahlung radiation from high-energy electrons colliding with nuclei.[17] Together with independent evidence, this led us and other researchers to believe that the hot spots contained ions and electrons at very high energies in the range of interest for advanced fuel fusion.[11, 18-19] However, most researchers in the field have believed that the hard radiation was generated not in the plasma itself but by the electron beam when it strikes the end of the anode, and that the hot spot plasma was relatively cool with electron energies of only several kiloelectron volts, too low for hydrogen-boron fusion.[18] This key question remained unresolved by previous experiments.

To clearly distinguish between the radiation from the anode end and that from the plasmoids, we used a cylindrical anode with 6-cm-deep recess at the pinch end (Fig.1). In this manner, the point where the electrons hit the anode, generating gamma rays, would be separated in space from the plasmoids, which are located at the plasma pinch very close to the end of the anode. We blocked the line of sight between the bottom of the hole and a set of X-ray detectors with a 5-cm-thick lead brick. This reduced the intensity of 1 Mev gamma rays by 50-fold. The line of sight from the plasmoids to the x-ray detectors, in contrast, passed through a 1-mm Be window, which allowed all but the lowest energy X-rays (below 1 KeV) to pass.

The experiments were performed at the Texas A&M University DPF facility that used a 268 microfarad capacitor bank, charged to either 30 or 35 kV. We used a shorter anode than is customary to increase the fill pressure which we anticipated on theoretical grounds would increase the plasmoid density and particle energy.[19] The cathode radius was 8.6 cm, anode radius 5.1 cm, anode length 24 cm, and insulator length 7.5 cm. A slotted knife edge around the insulator base was employed to enable reliable operation at the higher pressure. Shots were fired using fill gases of deuterium, helium or helium-deuterium mixtures at a 50-50 ratio by ion number. At 30 kV, optimal fill pressure was 12 torr for D, 14 torr for He-D and 16 torr for He. Peak current was typically 1.2- 1.3 MA at 30 kV and 1.4-1.5 MA at 35 kV.

To measure the energy of the X-rays and infer the electron energy, we placed 300-micron, 3-mm and 6-mm thick copper filters in front of the three X-ray detectors (NE102 scintillators with photomultiplier tubes (PMT)). The ratio of X-ray intensity at a given instant recorded behind the 6mm filter to that recorded behind the 300 micron filter provided a measure of average X-ray energy, since only the more energetic X-rays could penetrate the thicker filter. We calculated average electron energy $T_e$ by comparing the ratio of the 6 mm/300 micron-filtered signals observed to that calculated for bremsstrahlung emitted by Maxwellian distributions of electrons at different electron temperatures. We used the ratio of the 3-mm signal to the 300-micron signal to test if the energy distribution was Maxwellian or non-Maxwellian.[21] We used remote scintillators and PMTs at 9.0 and 17.4 meters that were shielded by 5-cm lead bricks to measure the gamma ray pulses from the anode end, and the neutrons produced by deuterium fusion.[22]

In all but one of the eight shots with highest signal-to-noise (S/N) ratio, average electron energy $T_e$ exceeded 100 keV and peak $T_e$ generally exceeded 150 keV. (Table 1) (Calibration uncertainties of $\pm$ 10% imply $\pm$ 20% errors in $T_e$.) A high $T_e$ is vital in hydrogen-boron fusion, since the ions must have still higher energy, and at the very high densities required for net energy production collisions with electrons with average $T_e$ below 100-125 keV will cool ions too rapidly. By comparison, $T_e$ in Tokamak Fusion Test Reactor experiments was, at most, 11.5 keV.

The good agreement between peak $T_e$ calculated from the ratio of 6mm/300 microns and that calculated from the ratio of 3 mm/300 microns shows that the results are consistent with a Maxwellian distribution. (Fig.1) However, they are also consistent within calibration errors with monoenergetic electrons. In this case, average $T_e$ would range from 180 keV to 300 keV rather than from 80 keV to 215 keV for Maxwellian distributions.

The lead shields distinguished the x-rays generated by the plasmoids from the gamma rays coming from the anode tip. This demonstrated that we were indeed measuring electron energy in electrons trapped in the plasmoids, not just in the electrons in the beam. As measured by the remote scintillators, the gamma ray pulse signals were between 60 and 180 times smaller than the x-ray peaks so they could not affect the x-ray results. (Fig.2) For many of the shots, including all the He shots, the gamma ray pulses were not observable at all with the shielded detectors. In addition, the gamma ray pulses did not resemble the x-ray pulses, being 4-6 times shorter in duration (FWHM) with rise times generally 15 times faster. Finally the fact that the X-ray pulses were not detectable when filtered by the lead brick confirms that the X-ray energy is concentrated below 600 keV, in contrast to the more than 1 MeV gamma rays from the anode.

These results show clearly that the high energy X-rays are emitted from high-energy electrons within the plasma. But do they come from the high density plasmoids or do they come from a larger volume at lower density?

We have determined the density of the plasma that produces the neutrons by the DD reaction. There is strong evidence that this is the same volume that traps the high energy electrons and produces the hard X-ray radiation. In 25% of the deuterium shots at optimum fill pressure (two out of eight), we have observed 14.1 MeV neutrons. These 14.1 MeV neutrons are produced by the fusion of deuterium nuclei with tritons produced by DD fusion reactions and then trapped in the high-density regions by strong magnetic fields. This was observed in one of the two deuterium shots with the best X-ray data (shot 81602), as well as in one shot (shot 80910) that was made before the X-ray detectors were functional. Shot 80910 also had the highest number of DD neutrons.

In these two shots, the ratio of the number of 14.1 MeV neutrons to the number of 2.45 MeV neutrons provides a direct measure of density. In DD plasma approximately 0.9 tritons are produced by the D+D->T+p reaction for every 2.45 MeV neutron produced by the D+D->3He+n reaction. Thus the ratio of 14.1 Mev neutrons to 2.45 MeV neutrons is R = 0.9F, where F is the fraction of tritons that undergo fusion. Since F = $<\sigma v>n\tau$, where $<\sigma v>$ is the average product of the fusion cross section and the triton Maxwellian velocity distribution, n is the deuteron plasma density and $\tau$ is the confinement time for the tritons, therefore n = $F/<\sigma v>\tau$.

In both shots, the FWHM length of the DT pulse, $\tau$, is 10 ns, which is real pulse width, being significantly longer than the minimum width set by instrumental response times. (The recovery time of the scintillator and PMT is <1 ns, but since data is taken every 2 ns, the minimum pulse width observable is 4ns and still well under the 10 ns recorded here.)[23] In shot 81602, F = $6.6\pm1\times10^{-3}$ and for shot 80910, F = $5.4\pm1.8\times10^{-4}$. The shot-to-shot variability is typical of DPF functioning without optimization of all electrode dimensions, which was not possible with this set of experiments. We calculated n on the assumption that the tritons did not slow down significantly from their initial 1.01 MeV energy during the confinement time, and that DT neutron production was terminated as the plasmoid density decayed.[24] On this assumption, n = $3.3\pm0.5\times10^{21}/cm^3$ for shot 81602 and $2.8\pm0.9\times10^{20}/cm^3$ for shot 80910. This assumes, conservatively, that all the T has been produced prior to the DT neutron pulse. This represents a peak density and we take average density to be half these values, $1.6\pm0.2\times10^{21}/cm^3$ for shot 81602 and $1.4\pm0.5\times10^{20}/cm^3$ for shot 80910.[25]

Measurements of X-ray emitted power and $T_e$ give us that $n^2V = 1.5\times10^{34}/cm^3$, V is the volume of the reacting region, essentially identical to the $1.6\times10^{34}/cm^3$ that can be derived from DD neutron emission, neutron pulse duration and ion energy $T_i$.[26] V therefore is $5.5\times10^{-9}$ $cm^3$, indicating a core radius of 6 microns and an overall plasmoid radius of 24 microns. This is comparable to the dimensions of hotspots observed with pinhole X-ray cameras in many previous plasma focus experiments.[6-8, 27]

For shot 81602, measurement of the emitted ion beam provides indirect confirmation of the density estimate. Measurements with a 10 cm diameter Rogowski coil located at 65 cm from the end of the anode showed that the ion beam had a peak current of 140 A and total ion number of $8.8\times10^{12}$. Since the number of ions in the plasmoids nV must equal or exceed the number of ions in the beam, these measurements can combine with $n^2V$ to set an upper limit on n of $1.7\times10^{21}/cm^3$, identical to our estimate from DT neutrons, implying that essentially all the ions are evacuated into the beam.

This observed equality of the ion number in the plasmoid and in the ion beam allows us to estimate the density of other shots that did not yield DT neutrons. We can do this by dividing the $n^2V$ measured in each shot through X-ray or neutron observations by the beam ion number. This analysis indicates that n ranges from 0.9-3.0x10$^{20}$/cm$^3$, with an average of 1.5x10$^{20}$/cm$^3$ for He shots and 2.5x10$^{20}$/cm$^3$ for D shots. These values are also consistent with the non-observation of DT neutrons in the other D shots. The factor of ten variations between best n and average n is typical of DPF functioning without extensive electrode optimization.

The high densities measured in this experiment are quite comparable to those measured by a variety of techniques in other DPF research[11-13] and are not in themselves surprising. Many research teams have observed the same 10$^{20}$-10$^{21}$ /cm$^3$ range with similar operating conditions. But we emphasize that the present experiments are the first in which $T_e$ >100keV are associated with such high plasma densities.

There is abundant evidence that the high-energy electrons that produce the hard X-rays do in fact originate in the same dense hot spots that produce the neutrons and that therefore $T_i$, $T_e$ and n are all high in same time and space. The X-ray pulses are closely correlated in time and duration with the neutron pulses, overlapping with them in all cases, and with the X-ray peak preceding the center of the neutron pulse by only 5-18 ns (average 11 ns). The DT pulse peak for shot 81602 is within 8ns of the X-ray pulse peak. The duration of the x-ray pulses and neutron pulses are closely correlated, with X-ray pulse duration always 0.4-0.55 that of the neutron pulse. In addition, from the neutron production, pulse duration, and $T_i$, the quantity $n^2V$ was calculated for each deuterium shot .The same quantity can be calculated from $T_e$ and X-ray power for the two D shots with high X-ray S/N. The average $n^2V$ derived from the neutron data is 1.7x10$^{34}$/cm$^3$, while that derived from the x-ray data is 2.2x10$^{34}$/cm$^3$, which are essentially identical. This would be an extraordinary coincidence if the neutrons and X-rays did not in fact originate from the same volume of plasma.

At these high densities, electron-electron collisional heating is rapid, so there cannot be a cool background plasma with a high-energy component--essentially all the electrons must be energetic. For 200keV electrons with a density of 1.4x10$^{20}$ the electron heating time is only 6 ns, considerably shorter than the typical 20-50 ns x-ray pulse duration, and for n=1.7x10$^{21}$ (shot 81602) the heating time is only .5 ns.

Not only is $T_e$ high in these plasmoids, but so is $T_i$, ranging from 45 to 210 keV. (These $T_i$ assume a Maxwellian distribution of ion energies. For monoenergetic ions, the range would be from 90 to 350 keV). [28] For comparison, the ideal $T_i$ for hydrogen-boron fusion is 600 keV. For $T_i$ =145keV, the reaction rate is one third of its maximum. The highest $T_i$ achieved by TFTR was 44 keV [4]

For shot 81602, $T_i$ is 55Kev and ion confinement time τ is 54 ns. On this basis, nτ for this shot is 9.2x10$^{13}$sec/cm$^3$ and nτ$T_i$ is 5x10$^{15}$keVsec/cm$^3$, a record value for any fusion experiment. Simultaneous X-ray observations show an average $T_e$ of 145KeV. For other shots, which have smaller n, but larger $\tau$ and $T_i$, $n\tau T_i$ is typically $1\times10^{15}$ keVsec/cm$^3$.

The high density-confinement-time product is important for fusion performance since it is directly proportional to the fraction of fusion fuel burned. For comparison the best $n\tau T_i$ achieved in the Tokamak Fusion Test Reactor was $5.5\times10^{14}$ keVsec/cm$^3$ with $T_i$ of 44 keV and $T_e$ of only 11.5 keV. [4] (Of course, total fusion energy produced was orders of magnitude less than TFTR in our much more modest experiment, as was the cost of the experiment.)

The $T_i$ and $T_e$ achieved in these shots already match those needed for hydrogen-boron fusion. For the He shots, the atomic mass is close to the average atomic mass of pB11. Theoretical studies have shown that collisions with the alpha particle products of pB11 fusion preferentially heat boron ions faster than electrons for $T_e > 125$ keV[17]. With an $n\tau$ of $2\times10^{15}$ sec/cm$^3$, a factor of 20 above that achieved here, a large gap between ion and electron temperature will be created, allowing fusion energy to heat the plasmoid faster than X-ray emission cools it. Since $n\tau$ increases in DPF as $I^2$, these conditions should be achievable with optimized electrodes for I of 3-5 MA.

These results show the potential of the DPF as a possible device for advanced fuel fusion. Further work will be required to increase the efficiency of energy transfer into the plasmoid, which is low in these shots. Such high efficiency has been achieved with other DPF experiments[29], but not with the high density or particle energy achieved here. However, optimization of electrode design, which we were not able to accomplish in this experiment, may solve this problem. Our results show that neutron yield increases rapidly with increased sheath run-down velocity. Therefore, a reduced anode radius, which increases the B field and run-down velocity, could increase efficiency to much higher levels. A combination of high fill pressure, small anode radius, and high rundown velocity should be able to simultaneously achieve high efficiency of energy transfer as well as the high density and particle energy needed for advanced fuel fusion.

21. To eliminate the possibility that radiation from off-axis sources passed through the chamber wall, in later shots we also blocked the sides of the detectors with lead bricks. However, results from earlier shots were very similar to those from later shots, so all data was used. Electrical noise from the pinch was suppressed with aluminum foil shielding, but interfered with the X-ray signals on some shots. Our analysis utilized only the shots with a S/N ratio of more than 8 to 1. Shots with lower S/N had higher $T_e$, so their exclusion does not affect our conclusions. In the D and D-He shots, confusion of X-ray and neutron signals was avoided since the neutrons reached the detectors 140ns after the x-ray signals and were relatively small due to the 1-2 mm thickness of the x-ray scintillators.

22. The remote scintillators were calibrated for neutron detection against absolutely calibrated silver activation neutron counters. Their sensitivity to gamma rays and x-rays was calculated based on data in the literature that related neutron to X-ray and gamma-ray sensitivity.[30-31] They were cross calibrated with the X-ray detectors (by measuring output from the same shot with all lead bricks removed), providing an *absolute* calibration of the X-ray detectors. The X-ray detectors were cross-calibrated with each other in several shots with the filters removed. Relative calibration measurements were consistent to within 10% and the accuracy of the absolute neutron calibrations, and thus of the absolute X-ray calibrations, is estimated to be better than 30%.

23. The very short pulse duration also implies a small spread in triton velocity relative to the axis of the DPF. (The lines of sight to the neutron detectors are perpendicular to that axis.) This is consistent with the model of the plasmoid developed by Bostick, Nardi and other researchers and confirmed by x-ray pinhole images[10] in which the plasmoid has an "apple core" structure aligned with the DPF axis. Tritons contained by the strong magnetic fields within the plasmoid encounter the highest density plasma, and thus the most chance for fusion reactions while traveling along the plasmoid axis, and much lower density while re-circulating through the outer region of the plasmoid.

24. Even at a density of $3.3 \times 10^{21}/cm^3$, the energy loss time for tritons in ion-ion collisions is 55 ns, much longer than $\tau$, supporting the assumption that the tritons have not slowed substantially by the end of the pulse. The energy loss time for the tritons at the same density in collisions with electrons at 100keV energy is longer still--350 ns.(A beam plasma interaction that slows the tritons collectively cannot be entirely ruled out, although we believe it unlikely. If the tritons are slowed down to low energies during the 10 ns pulse, the reaction rate could be increased by as much as 3.5, reducing the calculated density by the same factor.)

25. We have ruled out any source for the DT neutrons other than a dense plasmoid where the tritons are produced, along with the DD neutrons. Collisions of tritons with deuterated metal in either the chamber wall or the anode, even assuming (unrealistically) that there are one deuterium atom for every copper or iron atom, would produce a burn fraction F of $< 2.7 \times 10^{-5}$, a factor of 10 less than that observed for shot 80910 and a factor of 120 less than that observed for shot 81602. The tritons ejected in the ion beam will hit a copper target which is additionally shielded from the neutron detectors by a 60-cm thick concrete safety wall, further reducing any possible beam-target contribution to the DT neutrons. Nor can the DT detections possibly be cosmic rays or other noise sources. The delay time between the DT neutron peak at the 9 meter detector and that at the 17.4 meter detector is identical (with the 2 ns measurement window) to that for 14.1 MeV neutrons in shot 81602 and differs by only 4 ns (1%) for shot 80910. There are no other such pulses at unrelated times (except for the much earlier gamma ray pulse). While the number of neutrons actually detected is not large (7 for shot 81602 and 3 for shot 80910), even a single 14.1 MeV neutron produced a pulse four times the instrument noise level. The isotropic distribution of the neutrons is indicated by their detection at both detectors, which are 90 degrees apart in azimuthal direction.

26. $T_i$ and $\tau$ for the DD neutron production pulses were determined as follows. The FWHM was measured at 9 meters and at 17.4 meters. We made the simplifying assumption that the velocity distribution was Maxwellian. On that basis $FWHM^2 = \tau^2 + t^2 T_i/2E$, where t is the transit time for a neutron of energy E to reach the detector and $T_i$ is the ion temperature, averaged over the duration of the pulse. While it is unlikely the velocity distribution is in fact Maxwellian, monoenergetic ions would produce a different shaped neutron pulse than is observed, so a Maxwellian approximation is roughly accurate.

27. We did not in these experiments attempt to directly measure the source size of the high energy X-rays. There are difficulties with measuring the radius of the source of high-energy X-rays with a pinhole cameras is routine for low-energy x-rays. For a reasonable field of view, a pinhole must be cut in material that is thinner than the pinhole diameter. However, to have adequate contrast for X-ray energy above 100keV, even lead thickness must be >200 microns, so resolution of tens of microns cannot be obtained in this manner. (After our experiment, Castillo-Mejia et al, [32] using an image intensifier, demonstrated that the source size of hard x-rays in their smaller PDF was of the order of 10 microns, although in this experiment they did not distinguish between X-rays from the plasmoids and gamma rays from the electron beam collision with the anode.)

This research was funded by Jet Propulsion Laboratory and the Texas Engineering Experiment Station. The authors thank Paul Straight for his technical assistance.

**Correspondence should be addressed to E.L. (e-mail: elerner@igc.org).**

Table 1

| Fill gas | D | D-He | He |
|---|---|---|---|
| x-ray pulse duration , ns | 45 | 30 | 25 |
| Average, $T_e$ keV | 140-215 | 120-190 | 80-150 |
| Peak, $T_e$ keV | 210-500 | 180-250 | 100-210 |
| Neutronsx$10^9$ | 10-30 | 1-4 | |
| Neutron pulse duration, ns | 60 | 50 | |

Fig. 1 In a dense plasma focus, a small plasmoid or hotspot forms near the end of the anode, emitting an electron beam towards the anode, an ion beam away from the anode as well as x-rays and neutrons from fusion reactions. X-rays from the plasmoid pass through a 1 mm thick Be window to the x-ray detectors, but gamma rays from collisions of the electron beam with the anode are blocked by a 5 cm thick lead brick. Neutrons pass through apertures in a 60 cm concrete safety wall to neutrons detectors at 9 m and 17.4 m. Lines of sight of the two neutron detectors are 90 egress apart in azimuthal direction. A Rogowski coil at 65 cm from the anode measures the ion beam.

Fig. 2. X-ray power output(violet), electron average energy $T_e$ calculated from ratio of 6mm/300 micron -filtered output(red), $T_e$ from ratio of 3mm/330 micron filtered output (green) for a single D shot(35 kV, 15 torr, 9x$10^9$ neutrons). $T_e$ is in keV, while x-ray output, measured by 300 micron filtered detector, is in units of 100 W total emitted power. Time unit is 2ns. The close correspondence of the two measures of $T_e$ in the latter half of the pulse is consistent with a Maxwellian electron distribution, while the higher 6mm/300 micron $T_e$ in the earlier part of the pulse indicates a broader-than Maxwellian energy distribution. Average $T_e$ for this pulse is 200keV and $T_I$ derived from neutron time of flight measurement is 300keV. Note that x-ray pulse duration and overall X-ray energy emitted is considerably greater in this shot at 35kV, with peak current of 1.5MA than in Fig. 2, at 30kV and 1.3MA, showing rapid scaling of X-ray output.

Fig. 3 (a) dI/dt of device current(blue), x-ray output(violet), ion beam current (black), $T_e$ from 6 mm/300 micron ratio (red) , gamma-ray output through 5cm lead shield of remote PMT(green) from a single D shot((shot 81602, 30 kV, 12 torr, 1.0 x$10^{10}$ neutrons). Units :dI/dt 100 A/ns (inverted); x-ray output , kW; ion beam current, 5 A; , $T_e$ ,10 keV ; gamma ray output, 20 W of power through the shield. All traces are referred to time of radiation from the pinch region. The shielded gamma ray pulse is far smaller than the unshielded x-ray pulses and differs from it in shape and timing. The gamma ray pulse generated by the electron beam hitting the anode and the ion beam are clearly correlated as both are with the dI/dt pinch. Heating of the electrons in the plasmoid continues as the beam tail off, reaching a peak 10 ns later. Energy transferred to the ions leads to emission of neutrons(b) (kW of total neutron power). The broadening of the neutron peak due to the energy spread of the neutrons has been compensated for, although the peak is probably somewhat shorter in duration than portrayed here. In this shot, near the peak of neutron power, 14 MeV neutrons are emitted from collisions of a beam of trapped tritium with deuterium(c) (kW of DT neutron power). This occurs only 8 ns after the X-ray peak.

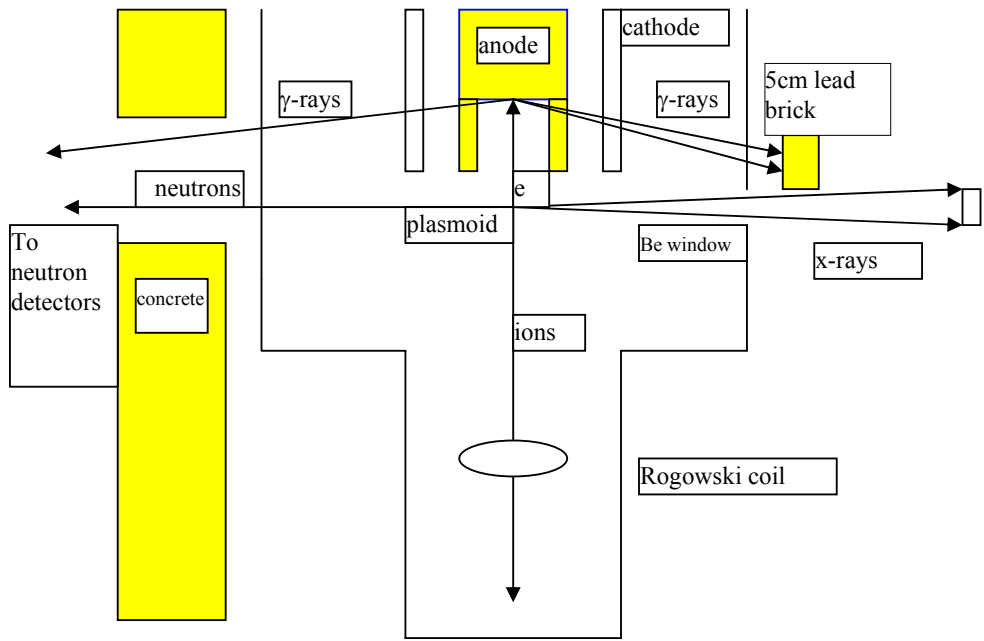

Fig. 1

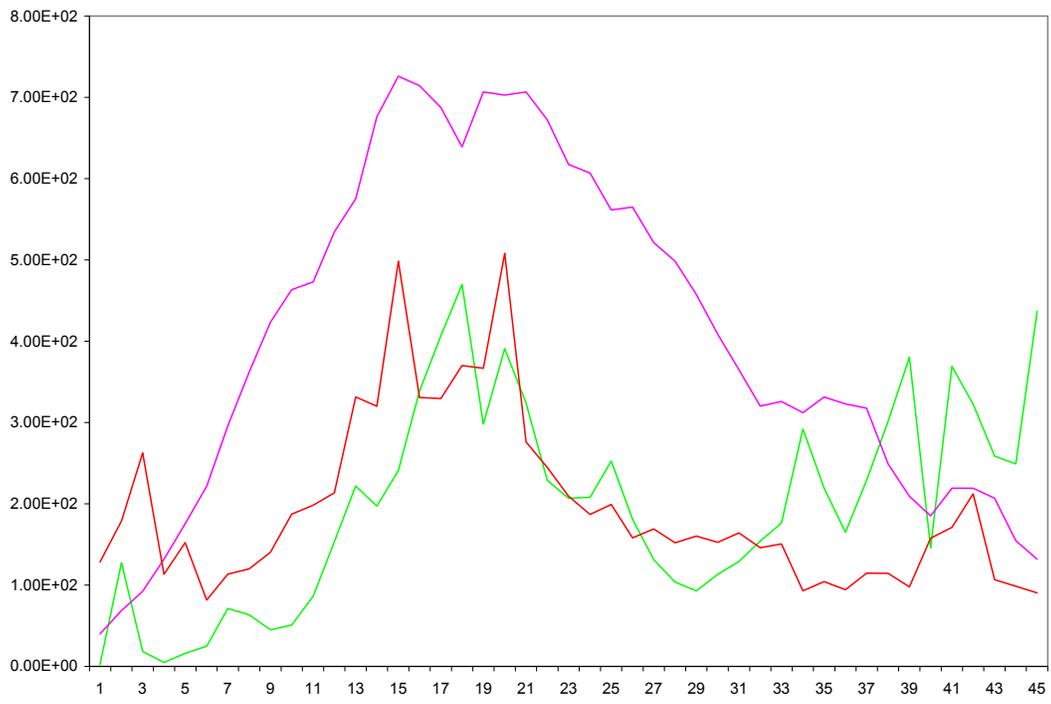

Fig. 2

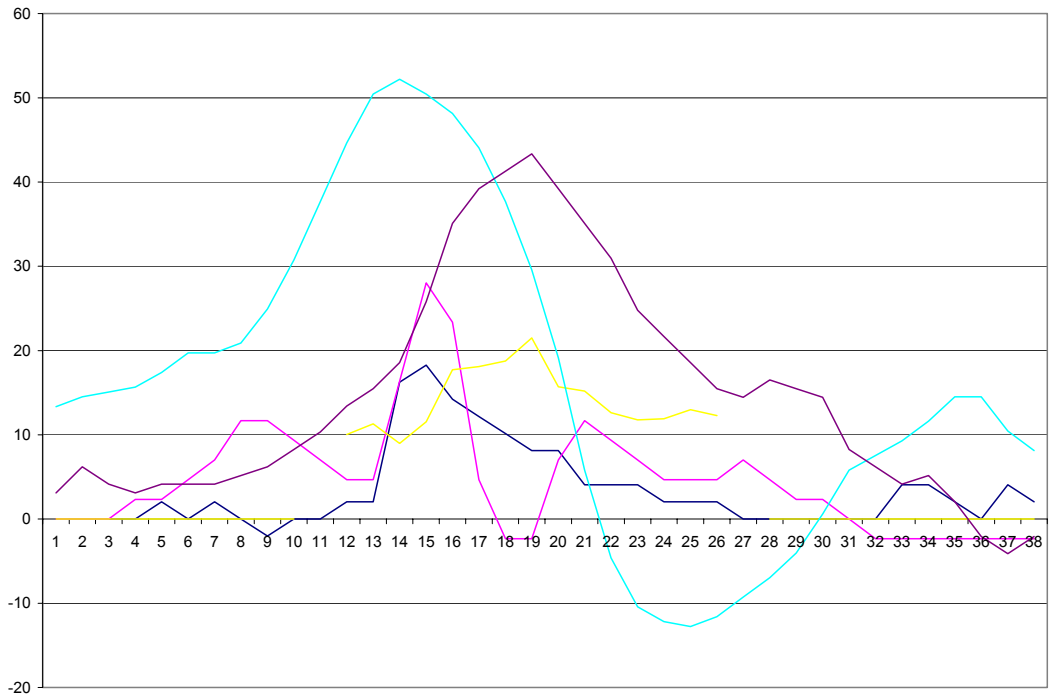

Fig. 3a

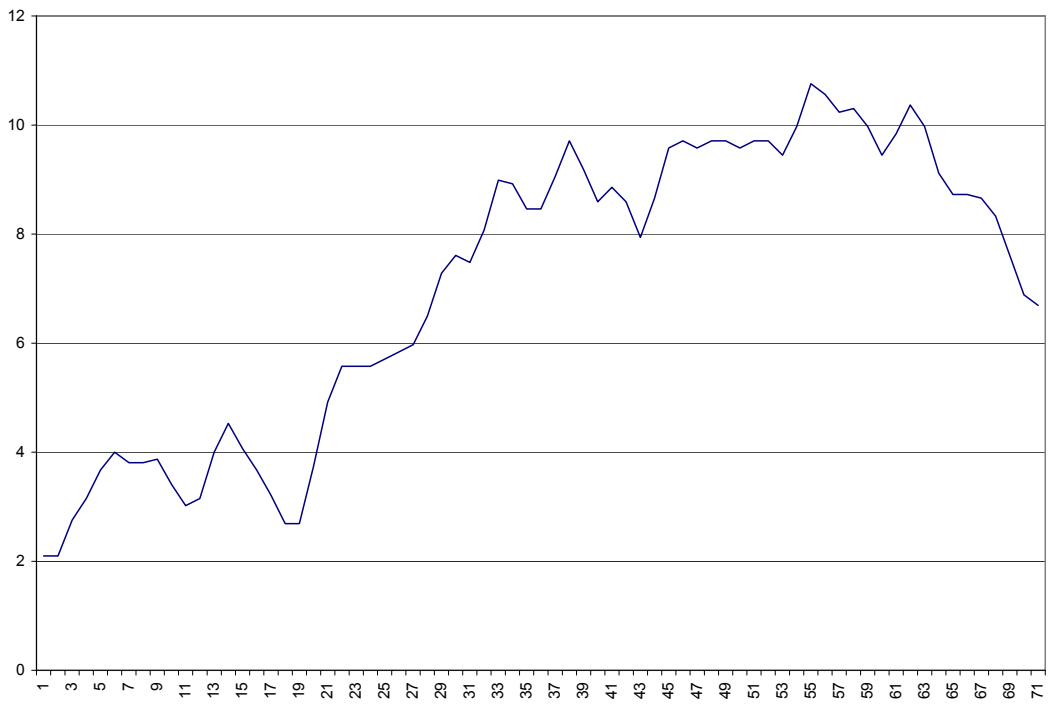

Fig. 3b

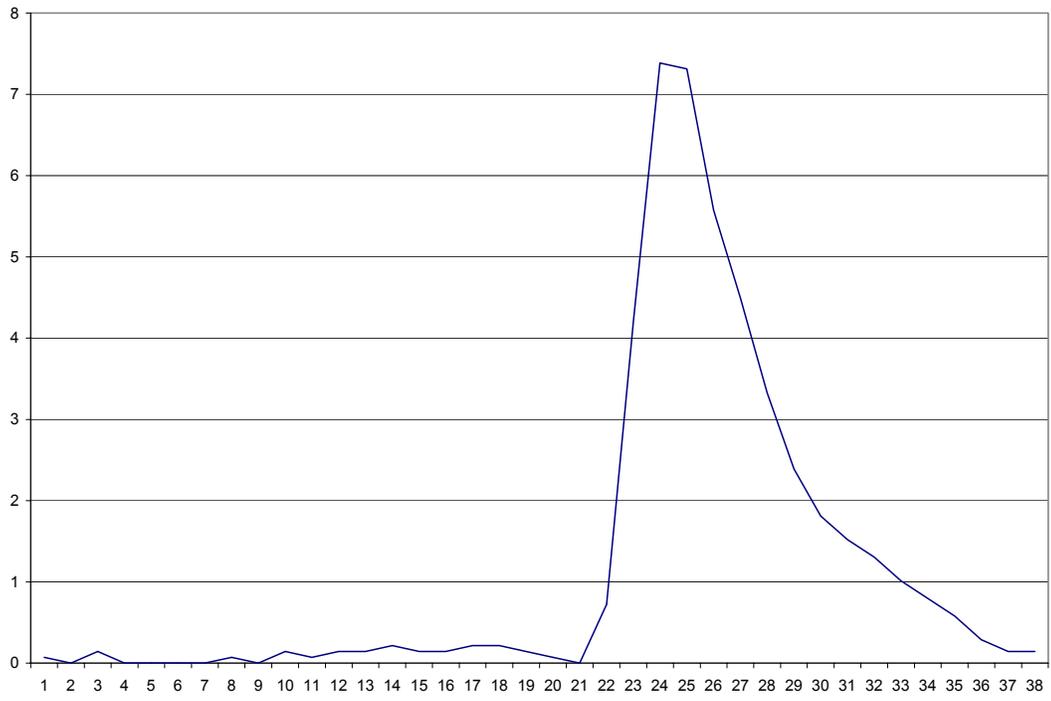

Fig. 3c